\documentstyle[prl,aps]{revtex}
\begin{document}
\draft
\title{\hfill{\rm {CERN-TH/2002-077}}\\
\hfill{\rm {hep-ph/0204051}}\\
\hfill{\rm {April, 2002}}\\
{\bf {The Problem of Large Leptonic Mixing}}}
\author{Joaquim I. Silva-Marcos\cite{juca}}
\address{Theoretical Physics Division, CERN, \\
CH-1211 Geneva 23, Switzerland}
\maketitle

\begin{abstract}
Unlike in the quark sector where simple $S_3$ permutation symmetries can
generate the general features of quark masses and mixings, we find it
impossible (under conditions of hierarchy for the charged leptons and 
without considering the see-saw mechanism or a more elaborate
extension of the SM) to guarantee large leptonic mixing angles with any
general symmetry or transformation of only known particles. If such
symmetries exist, they must be realized in more extended scenarios.
\end{abstract}

Recent neutrino data \cite{data} have not only provided clear evidence
pointing towards neutrino oscillations with very large mixing angles and
non-zero neutrino masses, but have also added more questions to one of the
most intriguing puzzles in particle physics: the flavour problem. We do not
know of any fundamental theory of flavour, but several specific patterns for
the fermion mass matrices have been proposed that account for the data. Our
hope is to find some pattern that may point towards the existence of some
family symmetry at a higher energy scale \cite{patterns}. For instance, in
the quark sector, an $S_{3L}^q\times S_{3R}^u\times S_{3R}^d$ family
permutation symmetry (acting on the left-handed quark doublets, the
right-handed up quarks and the right-handed down quarks) automatically leads
to quark mass matrices $M_u$, $M_d$ proportional to the so-called democratic
mass matrix \cite{demo}, which has all elements equal to unity. In the
democratic limit, only the third generation acquires mass and the
Cabibbo--Kobayashi--Maskawa (CKM) matrix is the unit matrix. This is a
remarkable result. Experimentally, one knows that there is a strong
hierarchy in the value of the quark masses. The first two generations of
quarks are much lighter than the third one, and the observed CKM matrix is
close to the unit matrix. When the permutation symmetry is broken, the first
two generations acquire non-vanishing masses, and a non-trivial CKM matrix
is generated.

Inspired by this, one may be tempted to try and find some
symmetry for the lepton sector. However, one is then confronted with the
problem of generating the large leptonic mixing. Let us assume, e.g. as in 
\cite{yanagi}, that the large mass difference between the neutrinos and the
other leptons comes from a Majorana Yukawa term in the Lagrangian: 
\begin{equation}
\label{effect}-{\cal L}=\frac{\lambda _{ij}^\nu }M\ \phi ^{\dagger }\phi
^{\dagger }L_iL_j+\lambda _{ij}^e\ \overline{L}_i\phi e_{jR}+{\rm h.c.}
\end{equation}
where the $L_i$ are the left-handed doublets, $\phi $ the Higgs field, $%
e_{jR}$ the right-handed charged lepton singlets and $M$ a large mass. As in
the case for the quarks, an $S_{3L}\times S_{3R}$ symmetry, acting on the
left-handed lepton doublets and the right-handed charged lepton singlets,
leads to a charged lepton mass matrix proportional to the 
democratic matrix\footnote{It must be enphasized that there 
exist many other 
symmetries (which are not permutations and) that require the charged
lepton mass matrix to be proportional to the democratic matrix,
$M_e=\lambda \Delta $, e.g. 
a $Z_3$ symmetry as proposed in Ref. \cite{z3}.},
denoted by $\Delta $. However, the most general mass matrix obtained from
the Majorana term in Eq. (\ref{effect}), and allowed by the 
symmetry, is of the form $\lambda \Delta +\mu {1\>\!\!\!{\rm I}}$. As was
pointed out \cite{yanagi}, there is no reason to expect that $\lambda $ and $%
\mu $ should not be of the same order of magnitude. As a result, both the
charged lepton mass matrix and the neutrino mass matrix are, in leading
order, diagonalized by the same unitary matrix, and the leptonic mixing
matrix will just be the unit matrix. It is clear, unless one puts $\lambda =0
$ by hand, that no large angles can be generated by a small breaking of the $%
S_{3L}\times S_{3R}$ symmetry. By making the ad-hoc assumption that the
coefficient $\lambda $ vanishes, one can, of course, obtain the required
large lepton mixing \cite{fritzsch}, but this is not dictated by the
symmetry. More precisely, the Lagrangian does not acquire any
new symmetry in the limit where $\lambda $ vanishes. Therefore, setting $%
\lambda =0$ clearly violates 't Hooft's naturalness principle \cite{hooft}.

In this Letter we shall prove that this problem for the leptons is, indeed,
much more general. One could imagine that some other symmetry (or
representation of the leptons) might exist that would require the charged
lepton mass matrix to be proportional to $\Delta $, while at the same
time preventing the neutrino Majorana mass from acquiring a similar term. We
shall prove that this is impossible. Consequently, it will also be
impossible to guarantee large mixing angles necessary to solve the
atmospheric neutrino problem\footnote{%
We shall work in a democratic weak basis. Obviously, our statement can be
extended to any other convenient basis.}.

Let us assume that the left-handed doublets and right-handed charged lepton
singlets transform under some general symmetry as 
\begin{equation}
\label{discr}
\begin{array}{l}
L_i\quad \rightarrow \quad P_{ij}\ L_j \\ 
e_{iR}\quad \rightarrow \quad Q_{ij}\ e_{jR} 
\end{array}
\quad 
\end{equation}
The charged lepton and neutrino mass matrices must then be invariant under 
\begin{equation}
\label{invar}
\begin{array}{lll}
P^{\dagger }\cdot M_e\cdot Q=M_e & \ ;\quad & P^T\cdot M_\nu \cdot P=M_\nu 
\end{array}
\end{equation}

If Eq. (\ref{invar}), for the charged leptons, is to impose\footnote{%
Our argument, here, requires that the relation for the charged leptons (i.e.
the symmetry) completely determines $M_e$ up to a multiplicative constant.
This means that the symmetry must be truly effective, and not simply,
require that $M_e$ be a rank-1 matrix. On the contrary, if the symmetry 
is not effective, i.e. if $M_e$ is not completely determined up to a 
multiplicative
constant as e.g. in the cases $P=Q={1\>\!\!\!{\rm I}}$, or 
$P={1\>\!\!\!{\rm %
I}}$, $Q={\rm diag}(-1,-1,1)$, or even $P={\rm diag}(i,1,1)$, $Q={\rm diag}%
(-1,-1,1)$, then our argument does not apply.} $M_e\sim \Delta $, then, in 
particular, it must be valid for $%
\Delta $ itself. Therefore, one must have $P^{\dagger }\cdot \Delta \cdot
Q=\Delta $. Squaring and using $\Delta ^2=3\Delta $, one finds $P^{\dagger
}\cdot \Delta \cdot P=\Delta $, which means that $P\cdot \Delta =\Delta
\cdot P$ and thus $P^T\cdot \Delta =\Delta \cdot P^T$. Using again Eq. (\ref
{invar}), one can now also find extra relations for $M_\nu $: 
\begin{equation}
\label{valid}
\begin{array}{c}
P^T\cdot \Delta M_\nu \cdot P=\Delta \cdot P^T\cdot M_\nu \cdot P=\Delta
M_\nu  \\ 
P^T\cdot M_\nu \Delta \cdot P=P^T\cdot M_\nu \cdot P\cdot \Delta =M_\nu
\Delta  \\ 
P^T\cdot \Delta M_\nu \Delta \cdot P=\Delta \cdot P^T\cdot M_\nu \cdot
P\cdot \Delta =\Delta M_\nu \Delta 
\end{array}
\end{equation}
Therefore, whatever the neutrino mass matrix is, nothing prevents it from
having additional large parts which can be written as: 
\begin{equation}
\label{contri}M_\nu +\lambda ^{\prime }\ \left( \Delta M_\nu +M_\nu \Delta
\right) +\lambda \ \Delta M_\nu \Delta 
\end{equation}
Note that, for any matrix $X$, one has $\Delta X\Delta =x\ \Delta $, where $%
x=\sum X_{ij}$. Thus, $M_\nu $ can have a large part proportional to $\Delta 
$. As a consequence, there exists no symmetry (be it discrete or not, but
realized in the form given in Eq. (\ref{discr})), that, as in Ref. \cite
{fritzsch}, would force the neutrino mass matrix to be strictly proportional
to ${1\>\!\!\!{\rm I}}$. It will also have a part proportional to $\Delta $.
In general, writing the neutrino mass matrix as $M_\nu =A+\lambda \Delta $,
where $A$ and $\lambda $ are of the same order, the symmetry cannot
guarantee the existence of large mixing angles necessary to solve the
atmospheric neutrino problem, because the term with $\Delta $ will be under
no restriction from the symmetry.

In fact, we can even be more precise and find very severe constraints for
the neutrino mass matrix and the mixing angles. From Eq. (\ref{invar}) for
the neutrino mass matrix one concludes that $M_\nu $ must also satisfy, $%
P^{\dagger }\cdot M_\nu ^{\dagger }M_\nu \cdot P=M_\nu ^{\dagger }M_\nu $.
Combining this with $P^{\dagger }\cdot \Delta \cdot Q=\Delta $, we find the
relation: 
\begin{equation}
\label{new}P^{\dagger }\cdot M_\nu ^{\dagger }M_\nu \Delta \cdot Q=M_\nu
^{\dagger }M_\nu \Delta 
\end{equation}
This seems to be an equation for the neutrino mass matrix, but, actually,
one must realize that it obeys the set of conditions that define the mass
matrix of the charged leptons as formulated in Eq. (\ref{invar}). Therefore, 
$M_\nu ^{\dagger }M_\nu \Delta $ must be proportional to the charged lepton
mass matrix i.e. to $\Delta $:

\begin{equation}
\label{new1}M_\nu ^{\dagger }M_\nu \Delta =p\Delta 
\end{equation}
Finally, we conclude that, because of the symmetry, the mass matrices of the
neutrinos and the charged leptons are intrinsically related. It is clear
that this has very strong consequences for the lepton mixing. It follows
that the matrix $F$ that diagonalizes $\Delta $ (i.e. the charged leptons on
the right) must also partially diagonalize $M_\nu ^{\dagger }M_\nu $. Using
Eq. (\ref{new1}), one finds that the matrix that diagonalizes $M_\nu
^{\dagger }M_\nu $ can be written as $F\cdot U$, where $U$ is a simple
unitary matrix with only significant elements in the $2\times 2$ sector,
i.e. $U_{13}=U_{23}=0$. As a result, the lepton mixing matrix will be just
this $U$. No perturbation, breaking the symmetry and giving small
contributions to $M_\nu $ and $M_e$, would be sufficient to obtain the large
mixing angles needed to solve the atmospheric neutrino problem\footnote{%
There could be one exception to this result: if the symmetry could force $%
M_\nu $ to be exactly proportional to a unitary matrix. Then, the
diagonalization of $M_\nu $ would not coincide with the diagonalization of $%
M_\nu ^{\dagger }M_\nu $, e.g. $M_\nu =(\omega -1){1\>\!\!\!{\rm I}}+\Delta $%
, where $\omega =e^{2\pi i/3}$. It would be possible to get large mixing
angles, because, in this specific case, small perturbations of the neutrino
mass matrix would cause singular effects\cite{jugugui}. However, one must
realize that unitarity can only be obtained with a very special combination
of terms, like setting $\lambda =0$ in our previous $S_{3L}\times S_{3R}$
symmetry example. As we have argued here, unitarity can never be forced by
the symmetry, because it is always possible to add any (small or large) term
proportional to $\Delta $ to $M_\nu $ and this would just lead to our small
mixing angles result.}.

One may try to avoid this difficulty and find less severe conditions, e.g.
by requiring that the charged leptons mass matrix be, not strictly
proportional to $\Delta $ but, only hierarchical: $M_e=V\Delta W$, where $V$
and $W$ are unitary matrices. But then, using analogous arguments, we find,
instead of Eq. (\ref{new1}), the relation $M_\nu ^{\dagger }M_\nu V\Delta
V^{\dagger }=pV\Delta V^{\dagger }$ between the neutrino mass matrix and the
charged lepton square mass matrix $H_e=M_eM_e^{\dagger }=3\ V\Delta
V^{\dagger }$. Clearly, this will just lead to the same result. It is
equivalent to an irrelevant change of weak basis. Another possibility would
be to state that neutrinos are of the Dirac type\footnote{%
For simplicity, we do not consider (heavy) mass terms for the right-handed
neutrinos and the see-saw mechanism. In fact, in the latter context, the
statement we make here is not necessarily true \cite{z3}.}. The relation for
the neutrino mass matrix in Eq. (\ref{invar}) will then read $P^{\dagger
}\cdot M_\nu \cdot R=M_\nu $, where $R$ is some transformation of the
right-handed neutrinos. We will then find that $M_\nu M_\nu ^{\dagger
}\Delta =p\Delta $. Again, we are unable to guarantee large mixing angles
from the symmetry or by a small breaking of it.

We find it, therefore, impossible (under the conditions stated with regard to 
the hierarchy 
and mass matrix of the charged leptons and without considering the 
see-saw
mechanism) to guarantee large mixing angles in the leptonic sector with any
general symmetry or transformation of only known leptons. If such symmetries
exist, they must be realized in more extended scenarios \cite{yanagida1}.
One may also interpret this result as a requirement to go beyond the SM and
it is indeed possible to construct a $Z_3$ permutation symmetry that, in the
context of the see-saw model, can lead to large mixing angles \cite{z3}.

\subsection*{Acknowledgements}

I am grateful to G.C. Branco for a thorough reading of the manuscript and
suggestions. This work received partial support from the Portuguese Ministry
of Science - Funda\c c\~ao para a Ci\^encia e Tecnologia and the CERN Theory
Division which I thank for the kind hospitality extended.


\begin{references}
\bibitem[*]{juca}  On leave from Instituto Superior T\'ecnico, Centro de
F\'\i sica das Interac\c c\~oes Fundamentais (CFIF), Lisbon, Portugal.
E-mail address: Joaquim.Silva-Marcos@cern.ch

\bibitem{data}  SuperKamiokande collaborations, Phys. Rev. Lett. 85 (2000)
3999, hep-ex/0106049, hep-ex/0103032, hep-ex/0103033; MACRO coll., Phys.
Lett. B 517 (2001) 59; GALLEX coll., Phys. Lett. B 447 (1999) 127; SAGE
coll., Phys Rev. C 60 (1999) 055801; GNO coll., Phys. Lett. B 490 (2000) 16;
SNO coll., Phys. Rev. Lett. 87 (2001) 71301.

\bibitem{patterns}  F. Vissani, hep-ph/9708483; E. Ma, Phys. Lett. B 442
(1998) 238; V. Barger, S. Pakvasa, T. J. Weiler and K. Whisnant, Phys. Lett.
B 437 (1998) 107; G. Altarelli and F. Feruglio, Phys. Rep. 320 (1999) 295;
Z. Berezhiani and A. Rossi, hep-ph/0003084; J.A. Casas, V. Di Clemente, A.
Ibarra and M. Quiros, hep-ph/9904295; J.I. Silva-Marcos, Phys. Rev. D 59
(1999) 091301; M. Jezabek and Y. Sumino, hep-ph/9904382; R. Barbieri, G. G.
Ross and A. Strumia, JHEP 10 (1999) 020; R. N. Mohapatra and S. Nussinov,
Phys. Rev. D 60 (1999) 013002; S. Lola and G. G. Ross, Nucl. Phys. B 553
(1999) 81; B. Stech, hep-ph/00006076; D. Black, A.H. Fariborz, S. Nasri and
J. Schechter, Phys. Rev. D 62 (2000) 073015; D. Falcone, Phys. Lett. B 475
(2000) 92; G. Altarelli, F. Feruglio and I. Masina, Phys. Lett. B 472 (2000)
382; G. Altarelli, hep-ph/0106085 and Nucl. Phys. Proc. Suppl. 87 (2000)
291; H. Fritzsch and Z. Xing, Phys. Rev. D 61 (2000) 073016; K. Fukuura, T.
Miura, E. Takasugi and M. Yoshimura, Phys. Rev. D 61 (2000) 073002; P.Q.
Hung, Phys. Rev. D 62 (2000) 053015; R. Adhikari, E. Ma and G. Rajasekaran,
Phys. Lett. B 486 (2000) 134; S.M. Barr and I. Dorsner, Phys. Rev. D 61
(2000) 033012; A. Kageyama, S. Kaneko, N. Shimoyama and M. Tanimoto,
hep-ph/0112359; E.Kh. Akhmedov, G.C. Branco, F.R. Joaquim and J.I.
Silva-Marcos, Phys. Lett. B 498 (2001) 237; H.J. Pan and G. Cheng,
hep-ph/0102060; K.R.S. Balaji, R.N. Mohapatra, M.K. Parida and E.A. Paschos,
Phys. Rev. D 63 (2001) 113002; N. Haba, J. Sato, M. Tanimoto and K.
Yoshioka, hep-ph/0101334; R. Kitano and, Y. Mimura, Phys. Rev. D 63 (2001)
016008; T. Miura, E. Takasugi and M. Yoshimura, Phys. Rev. D 63 (2001)
013001; F. Feruglio, A. Strumia and F. Vissani, hep-ph/0201291; M. Frigerio
and A.Yu. Smirnov, hep-ph/0202247; S. Lavignac, I. Masina and C.A. Savoy,
hep-ph/0202086.

\bibitem{demo}  H. Fritzsch, in Proc. of Europhys. Conf. on Flavour Mixing
in Weak Interactions (1984), Erice, Italy; G. C. Branco, M. N. Rebelo and J.
I. Silva-Marcos, Phys. Lett. B 237 (1990) 446; G. C. Branco and J. I.
Silva-Marcos, {Phys. Lett.} {B 359} (1995) 166; P.M. Fishbane and P. Kaus, {%
Phys. Rev.} {D 49} (1994) 3612, {Z. Phys.} {C 75} (1997) 1 and {Z. Phys.} {C 
}359 (1995) 166; G. C. Branco, D. Emmanuel-Costa and J. I. Silva-Marcos,
Phys. Rev. D 56 (1997) 107; P.M. Fisbane and P.Q. Hung, Phys. Rev. D 57
(1998) 2743; J.I. Silva-Marcos, Phys. Lett. B 443 (1998) 276 and
hep-ph/0102079; P.Q. Hung and M. Seco, hep-ph/0111013.

\bibitem{yanagi}  M. Fukugita, , M. Tanimoto and T. Yanagida, Phys. Rev. D
57 (1998) 4429.

\bibitem{fritzsch}  H. Fritzsch, Z.-Z. Xing, Phys. Lett. B 440 (1998) 313.

\bibitem{jugugui}  G. C. Branco, M. N. Rebelo and J. I. Silva-Marcos, Phys.
Rev. D 62 (2000) 073004 and Phys. Rev. Lett. 82 (1999) 683

\bibitem{hooft}  G. 't Hooft, {\it Naturalness, Chiral Symmetry and
Spontaneous Chiral Symmetry Breaking}, lecture given at Carg\`ese Summer
Institute 1979, p. 135.

\bibitem{yanagida1}  See as an example M. Tanimoto, T. Watari and T.
Yanagida, Phys. Lett. B 461 (1999) 345.

\bibitem{z3}  G. C. Branco and J. I. Silva-Marcos, Phys. Lett. B 526 (2002)
104.
\end{references}
\end{document}